\documentstyle[preprint,aps,floats,psfig]{revtex}

\begin{document}
\def\beq{\begin{equation}}
\def\eeq{\end{equation}}
\def\d{\delta}
\def\fourG{{{}^{(4)}G}}
\def\4R{{{}^{(4)}R}}
\def\H{{\cal H}}
\def\K{{\kappa}}
\def\mh{m_h^{}}
\def\vev#1{{\langle#1\rangle}}
\def\gev{{\rm GeV}}
\def\tev{{\rm TeV}}
\def\fbi{\rm fb^{-1}}
\def\lsim{\mathrel{\raise.3ex\hbox{$<$\kern-.75em\lower1ex\hbox{$\sim$}}}}
\def\gsim{\mathrel{\raise.3ex\hbox{$>$\kern-.75em\lower1ex\hbox{$\sim$}}}}

\newcommand{ \slashchar }[1]{\setbox0=\hbox{$#1$}   
   \dimen0=\wd0                                     
   \setbox1=\hbox{/} \dimen1=\wd1                   
   \ifdim\dimen0>\dimen1                            
      \rlap{\hbox to \dimen0{\hfil/\hfil}}          
      #1                                            
   \else                                            
            \rlap{\hbox to \dimen1{\hfil$#1$\hfil}}       
      /                                             
   \fi}                                             %

\tighten
\preprint{ \vbox{
\hbox{MADPH--00-1192}
\hbox{astro-ph/0009256}}}
\draft
\title{Supernova data may be unable to distinguish between quintessence 
and $k$-essence}
\author{V. Barger and Danny Marfatia\footnote{
 {\it {lastname}}@pheno.physics.wisc.edu }}
\vskip 0.3in
\address{Department of Physics, University of Wisconsin--Madison, WI 53706}
\vskip 0.1in

\maketitle

\begin{abstract}
{\rm We consider the efficacy of using luminosity distance measurements of 
deep redshift supernovae to discriminate between two forms of dark energy,
quintessence (a scalar field with canonical kinetic terms rolling down 
a potential) and $k$-essence (a scalar field whose cosmic evolution is 
driven entirely by non-linear kinetic terms). The primary phenomenological
distinction between the two types of models that 
can be quantified by supernova searches (at least in principle) is that the 
equation of state $w\equiv p/\rho$ of quintessence is falling today while 
that of $k$-essence is rising. By simulating $10^5$ possible datasets that
SNAP could obtain, we show that even if the mass density 
$\Omega_m$ is known exactly, 
 an ambiguity remains that may not allow a definitive 
distinction to be made between the two types of 
theories.
}
\end{abstract}
{{\bf Introduction.}}
Mounting evidence indicates that the universe is in an accelerating phase 
\cite{acc,perl1,riess,perl}. A phenomenological way to describe this is
 to postulate
the existence of energy (called dark energy) with 
negative pressure.
A popular form of dark energy that was
suggested a decade ago \cite{ratra} and has
received considerable attention recently is that of an evolving scalar field
that couples minimally to gravity \cite{quin}. It has become standard to call 
this field, quintessence \cite{dave,stein}. Since quintessence rolls down a 
potential, its equation of state is a function of redshift 
$p_Q(z)=w_Q(z)\rho_Q(z)$, where $-1\leq w_Q(0) < -1/3$
so as to explain the accelerated expansion of the universe. 
The cosmological constant is a special case corresponding to 
$w_Q(z)=-1$. However, quintessence models do not explain why
 the dark energy component dominates the universe only recently (the cosmic
coincidence problem).  The concept of $k$-inflation \cite{kinflation} when applied 
to quintessence \cite{chiba,k} addresses this problem; 
$k$-essence \cite{k} is a scalar field whose action has 
non-linear kinetic terms 
and no potential term. The non-linear kinetic terms lead
to dynamical attractor behavior that serves to avoid the cosmic coincidence
problem. This class of models preserves the welcome feature of tracker 
quintessence
models \cite{stein} in that cosmic evolution is insensitive to initial
conditions. Indubitably, it is important to distinguish
 between different models
of quintessence as characterized by their potentials, but it is 
even more urgent to discriminate
between different kinds of theories (like quintessence and $k$-essence) 
that seek to explain the acceleration of the universe.

Current luminosity distance $(d_L)$ measurements of type Ia supernovae 
(SNe Ia) at
 redshift $z\lsim1$ \cite{perl1,riess,perl}, 
though providing direct evidence of
cosmic acceleration, are unable to establish with a reasonable
 degree of certitude 
whether a cosmological constant or an evolving scalar field is responsible
for this acceleration. Numerous studies using supernova data 
(real or simulated) have considered the 
feasibility of determining the equation of state  and
 reconstructing the potential assuming that
the dark energy is quintessence \cite{es,maor,fitt1,fitt2,alb}.  
With the possibility \cite{yun} 
of obtaining a large dataset of supernova luminosity distance measurements to
deep redshifts from a satellite such as SNAP \cite{snap} or a dedicated
 dark matter telescope \cite{DMT}, this line of
investigation becomes all the more pertinent. There have been
conflicting interpretations of how useful SNAP-data would be in determining
$w_Q(z)$. It was emphasized in \cite{maor} that due to the lack of precision
in existing measurements of the current mass density 
$\Omega_m$, the value of $w_Q(0)$ would be uncertain
to 40$\%$ and absolutely nothing could be learned about the redshift
dependence of $w_Q$. 
Focusing instead on the rapid improvement expected in the observational 
determination of 
$\Omega_m$, the authors of \cite{fitt2,alb} show that various models of
quintessence can be distinguished provided $\Omega_m$ is 
sufficiently accurately measured.

In this Letter we take a broader view and allow for the possibility of 
dark energy other than quintessence. 
Throughout, we assume the universe to be flat as predicted by inflation 
and supported by measurements of the cosmic microwave background \cite{cmb}. 
We demonstrate that even if $\Omega_m$
is precisely known, it might still be possible to mistake
$k$-essence for quintessence and vice-versa. The
 principal feature that would enable supernova data to
distinguish between the two theories is that $k$-essence generically 
has $dw_k/dz<0${\footnote{Henceforth, we use the subscripts 
$k$ and $Q$ for  $k$-essence and quintessence, respectively.}}, whereas for 
most quintessence models $dw_Q/dz>0$ . 
To fit the supernova data, a choice of a
model-independent fitting function for the relative 
magnitude $m(z)$ is required, and we prefer the most 
physically transparent function derived using the expansion 
$w(z)=w_0+w_1 z+\ldots$, introduced in Refs.~\cite{maor,alb}. 
We analyze simulated datasets that SNAP could obtain in the case of the two 
theories for different values of $w_1\equiv dw/dz$ and show that the best fit could
provide misleading indications about the type of theory. 

Within the context of the two classes of theories under consideration, we would like to 
draw model-independent
conclusions. Consequently, we do not assume the existence of a prototypical or natural model
of quintessence or $k$-essence, but instead let data be the arbiter. 
In this approach we must sacrifice some 
predictivity for the sake of model-independence.
 Consider for example, the quintessence model defined by the potential
$V(Q) \propto 1/Q^{\alpha}$ \cite{ratra}. The equation of state $w_{Q}(z)$ is then a function of only one
parameter, $\alpha$, and by making use of say, the linear expansion $w(z)=w_0+w_1 z$, we are
effectively requiring the data to constrain one
additional parameter, which is clearly more difficult. 
\\

{{\bf Quintessence vs $k$-essence.}}
 In quintessence models the equation 
of state is given by
\beq
 w_Q(z)={p_Q(z) \over \rho_Q(z)}={{{1 \over 2}\dot{Q}^2-V(Q)} \over 
{{1 \over 2}\dot{Q}^2+V(Q)}}\,.
\eeq 
As the universe evolves, 
$V$ dominates over the slowly-varying $\dot{Q}^2$ and eventually
 $\dot{Q}^2 \ll V(Q)$ leads to $w_Q(0)\sim-1$. 
In most theoretically well-motivated models the equation of state of 
quintessence tracks that of matter in the matter-dominated epoch 
\cite{stein,theor} and only recently falls from $w_Q\sim0$ to $-1$
({\it i.e.} $dw_Q/dz>0$), thus triggering the accelerating phase. 

The evolution of the universe is more involved in the case of $k$-essence 
\cite{k},
and we briefly describe its features relevant to our study; $k$-essence
is a scalar field $\phi$ that (like quintessence) is minimally coupled 
to gravity, but has only (non-linear) kinetic terms in its action,
\begin{equation}
S_k=\int d^4 x~ \sqrt{-g} \{ 
{R\over 2\, \kappa^2}+{\tilde{p}(X) \over \phi^2}\}\,,
\label{action}
\end{equation}
where $\kappa^2=8\,\pi\, G$ and $X=\dot{\phi}^2/2$. The pressure of
$k$-essence is 
identical to the Lagrangian, \mbox{$p_k(z) = \tilde{p}(X)/\phi^2 $},
and its energy density is given by
\mbox{$\rho_k(z)=(2\,X\,\partial{\tilde{p}(X)}
/\partial{X}-\tilde{p}(X))/\phi^2$}.
Note that $2\,X\,\partial_X\tilde{p}>\tilde{p}$ is required for 
$\rho_k$ to be positive definite.
Because the kinetic term is non-canonical, the speed of sound 
$c_s^2=\partial_X p_k/\partial_X \rho_k$ is no longer
unity and a constraint has to be placed to ensure that it be real. 

Actions of the above type are standard
in string and supergravity theories. Usually the non-linear
terms become negligible when  attractor 
solutions that do not permit their damping are absent. When this is not the 
case, the non-linear dynamics can have profound consequences.
The most pleasing aspect of Eq.~(\ref{action}) is the dynamical attractor 
behavior emerging from it. Because of this property 
equipartition initial conditions are easily accommodated.
Moreover, $k$-essence can be made to transit from one attractor to another
in a manner that mimics cosmic evolution.
 If the attractor solutions of 
$k$-essence track the dominant energy component they
are called trackers (the radiation and dust trackers). 
There are two types of non-tracking attractors corresponding to $\Omega_k$
being close to $0$ or $1$. A ``de Sitter'' attractor is one with
$\Omega_k\sim 0$ and $w_k\sim -1$ while a ``$k$-attractor'' has 
$\Omega_k\sim 1$ and $w_k\gsim -1$.   
A realistic model of cosmic evolution would be one in which $k$-essence has 
a radiation tracker solution which 
may or may not be followed by a dust tracker solution, but must necessarily
reach either very close to a de Sitter attractor or have a $k$-attractor 
solution recently. It has been shown that if a dust tracker exists, a 
$k$-attractor cannot. It is still possible to generate an accelerating phase
(by approaching a de Sitter attractor)
{\it before} $k$-essence approaches the dust tracker, in the so-called 
``late dust tracker'' scenario. However, to achieve this a fine-tuning is 
required making it less natural. If a dust tracker does not exist, after the
radiation-dominated epoch, the $k$-field approaches the de Sitter attractor
(during which $\Omega_k\sim 0$), and after $\Omega_k$ increases sufficiently,
it moves to the $k$-attractor. This triggers cosmic acceleration. Let us call 
this the ``$k$-attractor'' scenario. This model
is natural and elegant. It is crucial to note that in both cases, 
$w_k$ rises from $-1$ to its value today; from
$w_k\sim -1$ close to the de Sitter attractor to a larger value as
  $k$-essence makes its way either to the dust tracker or the $k$-attractor.
Thus, in $k$-essence models it is generic that $dw_k/dz<0$.

As far as luminosity distance data from SNe Ia is concerned, the only 
distinguishing feature between quintessence and $k$-essence is that for
quintessence $dw_Q/dz>0$, and for $k$-essence $dw_k/dz<0$. With this in mind
we model the equations of state of quintessence and $k$-essence with the
expansions
\beq
w_Q(z)={w_Q}_0+{w_Q}_1\,z\,, \ \ \ \ \ {w_Q}_1>0
\eeq
\beq
w_k(z)={w_k}_0+{w_k}_1\,z\,, \ \ \ \ \ {w_k}_1<0\,,
\eeq
respectively. In principle there is no reason to terminate the expansions
at linear order, but by doing so we are implicitly considering 
models with the
least degeneracy while still permitting a redshift-dependent $w$. We find
that even this oversimplification proves insufficient to alleviate the
strong parameter correlations that make it difficult to tell the
two theories apart.
\\

{{\bf Modeling the luminosity distance.}} 
The distance modulus is related to the luminosity distance $d_L$ as
\beq
m(z)-M=5\,{\rm{log}} d_L +25\,,
\eeq  
where 
\beq
d_L(z)=c\,(1+z)\int_{0}^{z} {dz' \over H(z')}
\eeq  
and $m$ and $M$ are the apparent and absolute magnitudes of the source,
respectively. Because of the plenitude
of possible models that need comparison to a given $m(z)$ dataset, it is 
useful to adopt an appropriate fitting function in terms of
whose parameters the dataset may be described without recourse to a 
specific theory. Several fitting functions for $d_L$ have been
suggested and applied to existing and simulated data  
\cite{maor,fitt1,fitt2,alb}. As shown in \cite{alb} the fitting function
with the best fit (among the aforementioned suggestions) 
to existing models happens to be the one that lends
its parameters to the clearest physical interpretation. It is derived by 
expanding $w$ in a power series in $z$ \cite{maor,alb},
\beq
w(z)= w_0+w_1\,z+w_2\,z^2+\ldots\,.
\eeq
 Since we assume the universe to be flat, we have, 
in terms of the one additional parameter $\Omega_m$,
\begin{equation}
H_0\,d_L(z)=c\,(1+z)\int_{0}^{z} dz'\, [\Omega_m\,(1+z')^3+(1-\Omega_m)\,
e^{3\,w_1\,z'}\,(1+z')^{3\,(1+w_0-w_1)}]^{-1/2}\,,
\label{fit}
\end{equation}
where we have retained terms in $w(z)$ to linear order 
in the power expansion. 
Then the parameters of the fitting function and those of the theories under
consideration are in a one-to-one correspondence. The Hubble constant $H_0$ 
on the right will not
participate in fitting to data because a corresponding term in the 
fiducial model cancels it out. Note that $w_0=w(0)$ defines
the equation of state of dark energy today, and $w_1=dw/dz$ is the
quantity whose sign will help in discriminating between quintessence 
and $k$-essence.  
\\

{{\bf Simulating SNAP data.}}
 We construct $10^5$ possible datasets that SNAP
\cite{snap} could record. Each dataset contains 1915 supernovae with 
50, 1800, 50 and 15 supernovae in the redshift intervals (0,0.2), (0.2,1.2), 
(1.2,1.4) and (1.4,1.7), respectively. We bin with  50 supernovae
except for the last high redshift bin which consists of 15 observations. The
statistical error enters in the peak-brightness uncertainty and 
is assumed to be 0.15 mag per supernova. We conservatively 
neglect systematic errors.
All the theoretical models we consider have $\Omega_m=0.3$ and $w(0)=-0.7$.
 We will present our results as deviations from a fiducial model defined
by $\Omega_m=0.3$ and $w(z)=-0.7$ thus emphasizing the effect of a non-zero
$w_1$. 
 In the $\chi^2$-analysis we suppose that 
$\Omega_m$ has been measured exactly to be $0.3$.
In doing so, we eliminate one parameter in the fitting function (\ref{fit}) and
reduce the space of degenerate models considerably. 
Moreover, now {\it all} the degeneracy resides in the parameters of the 
dark energy theory; no degeneracy arises from the lack of precision
in the measurement of $\Omega_m$. Despite the truncation of $w(z)$ to
linear order, the neglect of systematic 
errors, and the elimination in degeneracy arising from an imprecisely 
known $\Omega_m$, we will find that it will not be possible to establish
that dark energy is either quintessence or $k$-essence with much conviction.
 If we relax these assumptions the situation gets worse. 

\begin{figure}[t]
\centerline{\mbox{\psfig{file=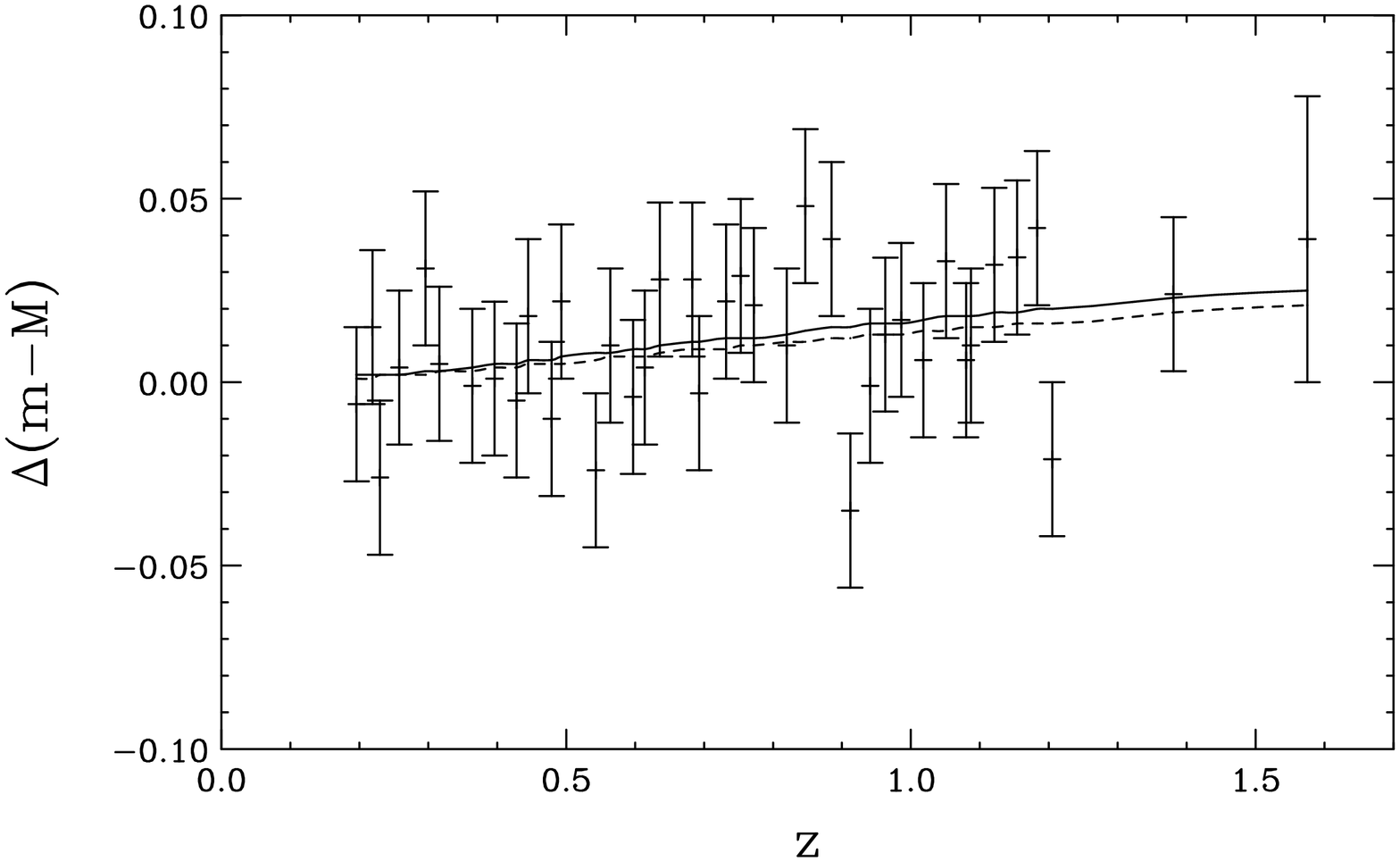,height=6cm,angle=90}
\psfig{file=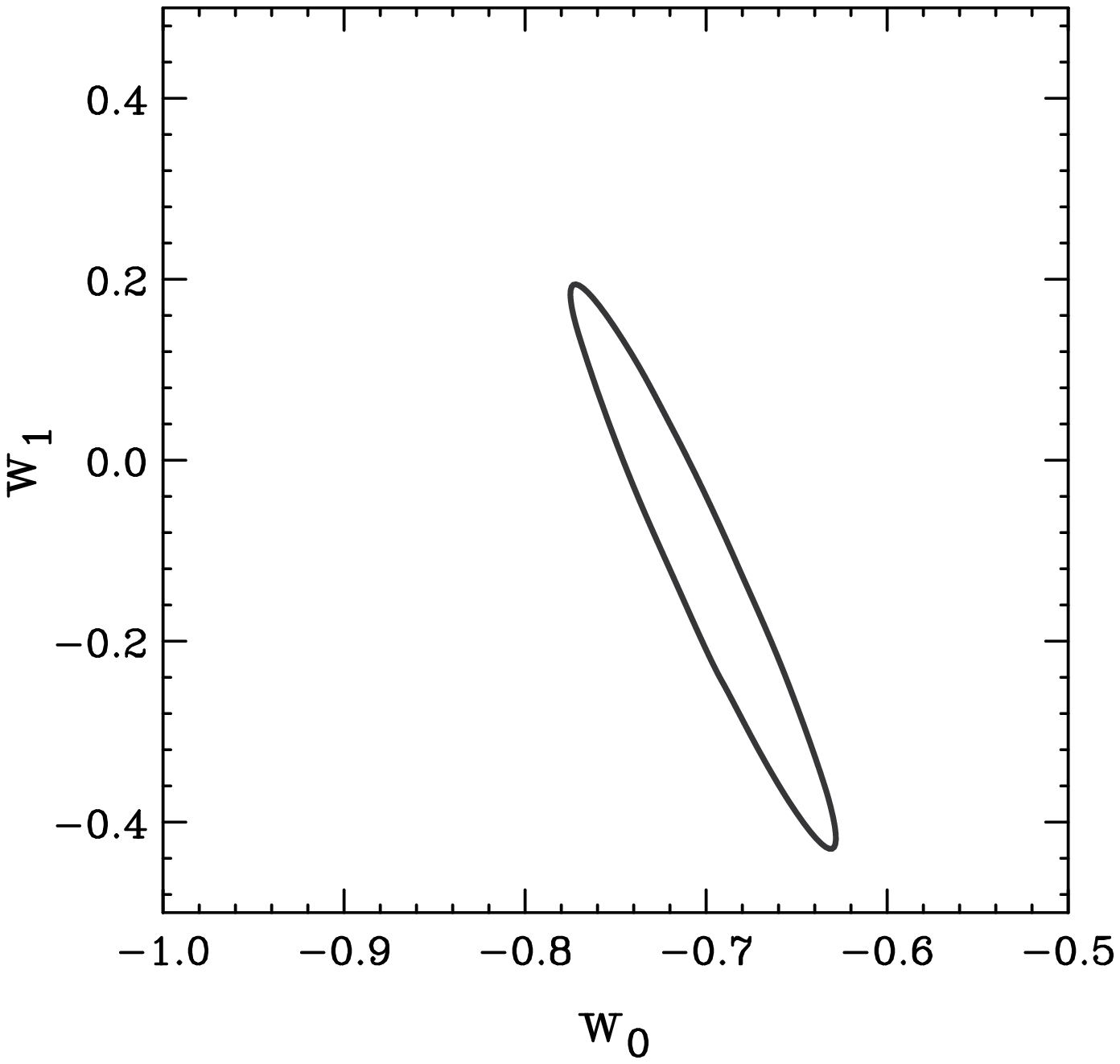,height=5.95cm,angle=90}}}
\bigskip
\caption[]{The difference in distance modulus between a $k$-essence 
model defined by $(\Omega_m , {w_k}_0 , {w_k}_1)=(0.3,-0.7,-0.1)$ 
 and the fiducial model $\Omega_m=0.3$,
$w(z)=-0.7$.  The
dashed line is the theoretical prediction. 
The solid line 
is the best fit to the simulated data with $({w}_0 , {w}_1)=(-0.7,-0.12)$,
demonstrating excellent agreement with the input theory. 
$\chi^2$ is minimized 
with the assumption that $\Omega_m=0.3$ precisely. The 
$w_0-w_1$ plane shows the $\Delta \chi^2=6.17$ (95.4\%
confidence level) contour. }

\label{sne3}
\end{figure}

\begin{figure}[t]
\centerline{\mbox{\psfig{file=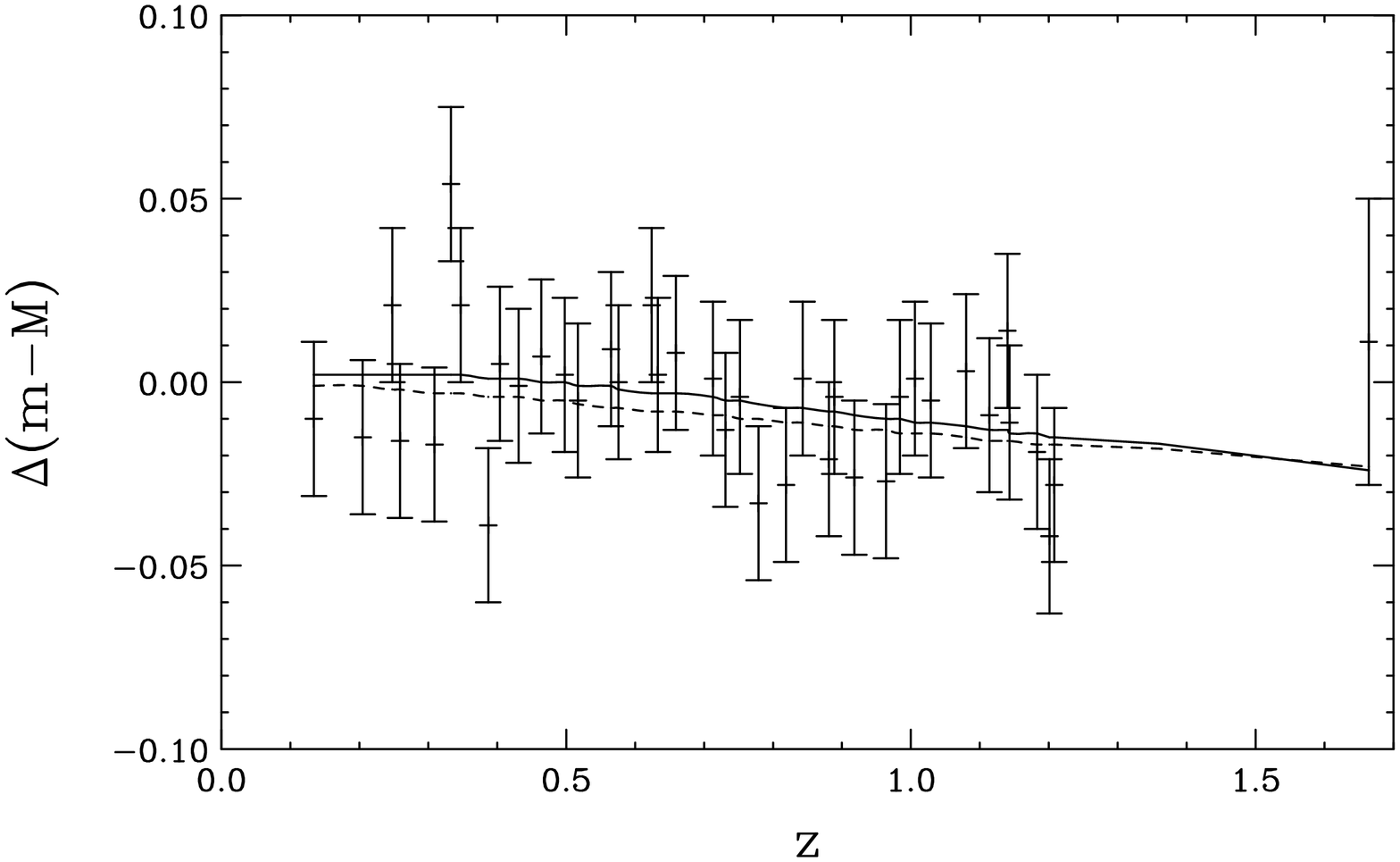,height=6cm,angle=90}
\psfig{file=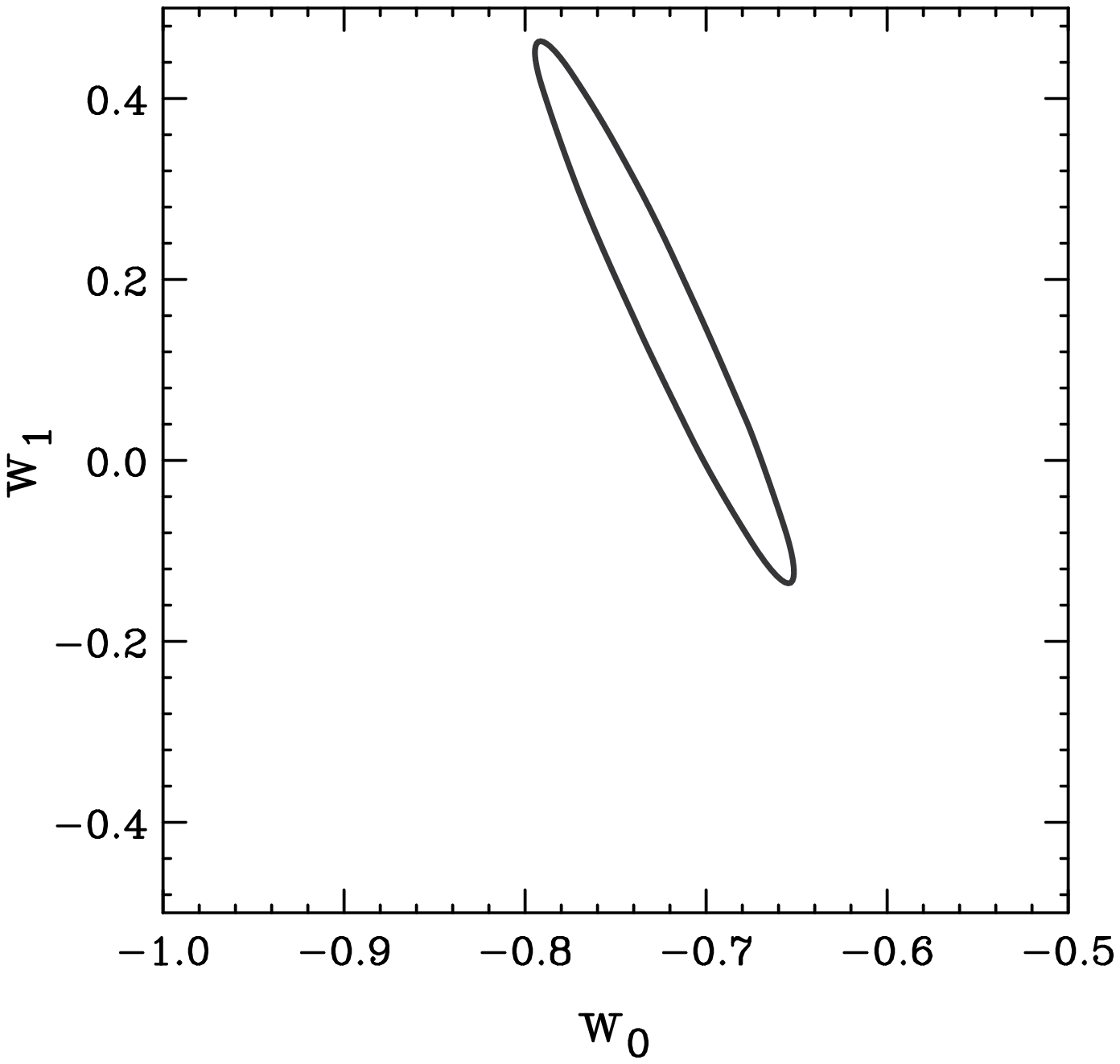,height=5.95cm,angle=90}}}
\bigskip
\caption[]{The difference in distance modulus between a quintessence
model defined by  $(\Omega_m , {w_Q}_0 , {w_Q}_1)=(0.3,-0.7,0.1)$ 
 and the fiducial model. 
The solid line 
is the best fit (fixing $\Omega_m=0.3$) to the simulated data with $({w}_0 , {w}_1)=(-0.72,0.16)$,
demonstrating convincing agreement with the input theory. The
$w_0-w_1$ plane shows the $\Delta \chi^2=6.17$ (95.4\%
confidence level) contour. }
\label{sne4}
\end{figure}

In Figs.~\ref{sne3} and \ref{sne4} we display examples of datasets for which 
the best fit curves agree remarkably well with that of the input theory. The 
difference in distance modulus is plotted relative to our fiducial model. 
Figure \ref{sne3} illustrates a $k$-essence
model with $(\Omega_m , {w_k}_0 , {w_k}_1)=(0.3,-0.7,-0.1)$. A $\chi^2$
analysis with $\Omega_m$ held fixed at 0.3 gives the best-fit 
parameters $({w}_0 , {w}_1)=(-0.7,-0.12)$ which agrees very well with the 
theoretical expectation. The $w_0-w_1$ plane shows the 
$\Delta \chi^2=6.17$ contour corresponding to a 95.4\% 
confidence level. The theoretical parameters are almost at the center of 
the ellipse.  In Fig.~\ref{sne4} we make similar plots for a quintessence
model with parameters  $(\Omega_m , {w_Q}_0 , {w_Q}_1)=(0.3,-0.7,0.1)$.
For this dataset $\chi^2$ is minimum at $({w}_0 , {w}_1)=(-0.72,0.16)$ where
 we have not allowed $\Omega_m$ to vary.


\begin{figure}[t]
\centerline{\mbox{\psfig{file=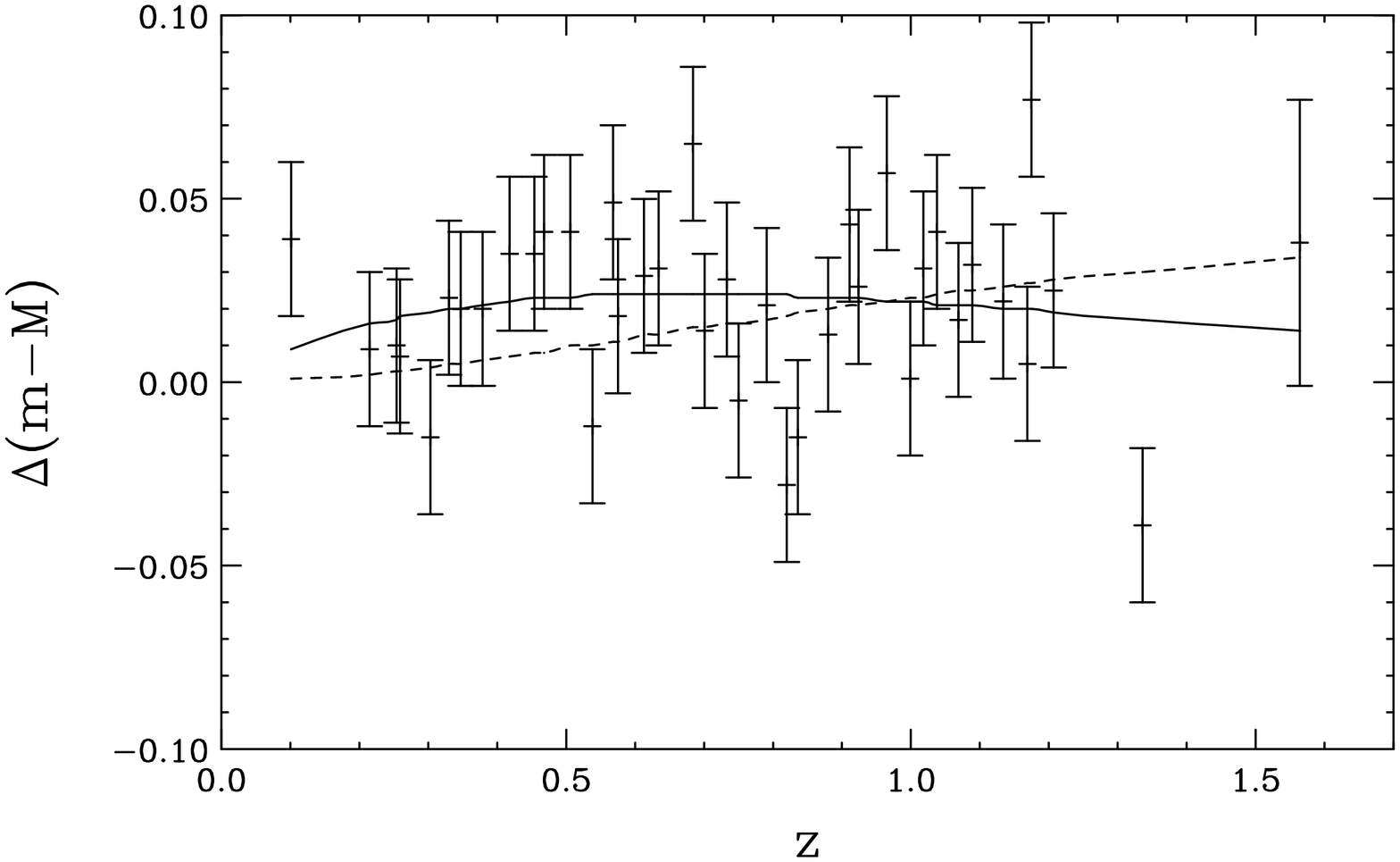,height=6cm,angle=90}
\psfig{file=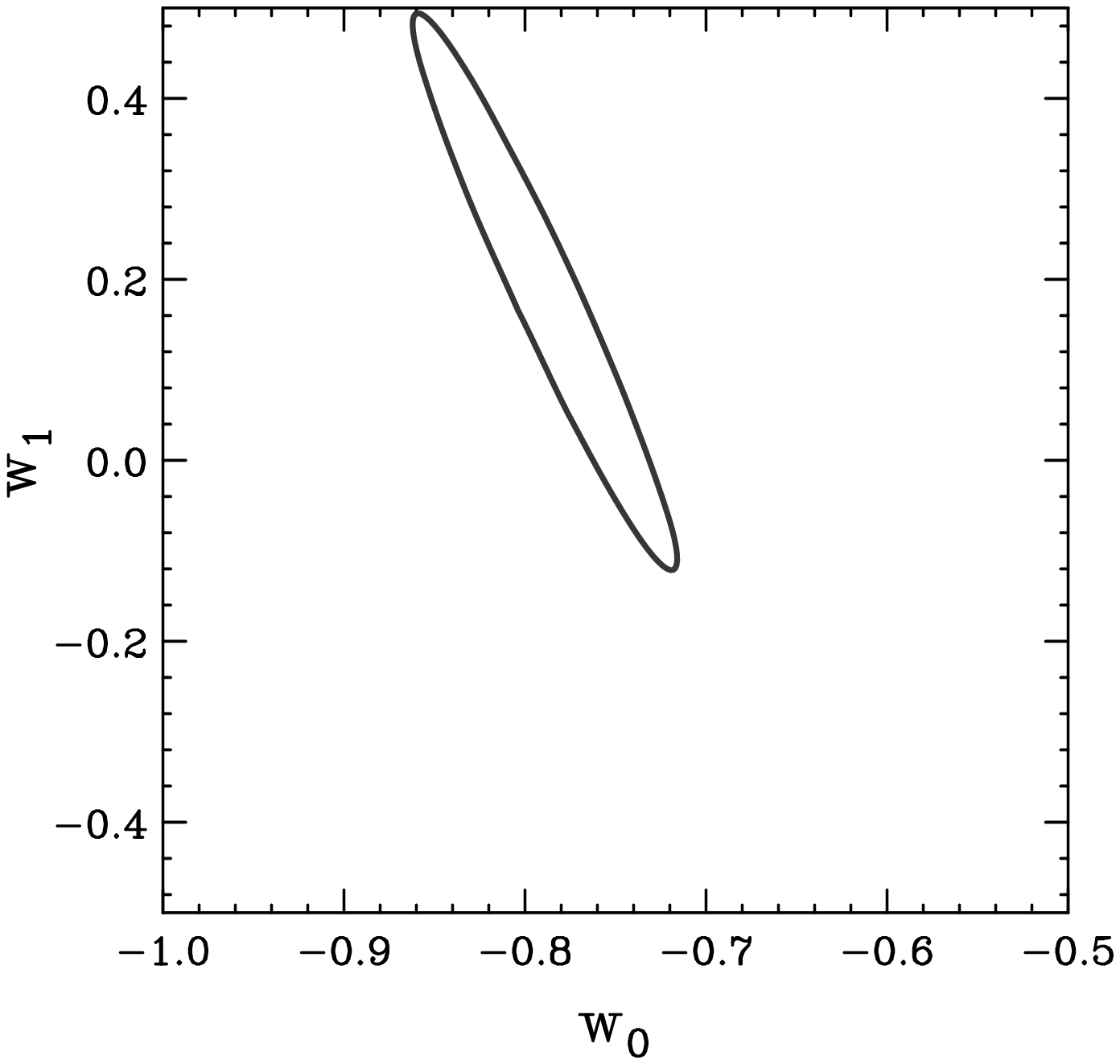,height=5.95cm,angle=90}}}
\bigskip
\caption[]{The difference in distance modulus between a $k$-essence 
model defined by $(\Omega_m , {w_k}_0 , {w_k}_1)=(0.3,-0.7,-0.17)$ 
and the fiducial model.   
The best fit to the simulated data is $({w}_0 , {w}_1)=(-0.79,0.19)$,
matching a quintessence model. $\chi^2$ is minimized 
with the assumption that $\Omega_m=0.3$ precisely. The 
$w_0-w_1$ plane shows that the $\Delta \chi^2=6.17$ (95.4\%
confidence level) contour lies mainly in the upper half-plane
and does not include the theoretical model.   }
\label{sne1}
\end{figure}

\begin{figure}[t]
\centerline{\mbox{\psfig{file=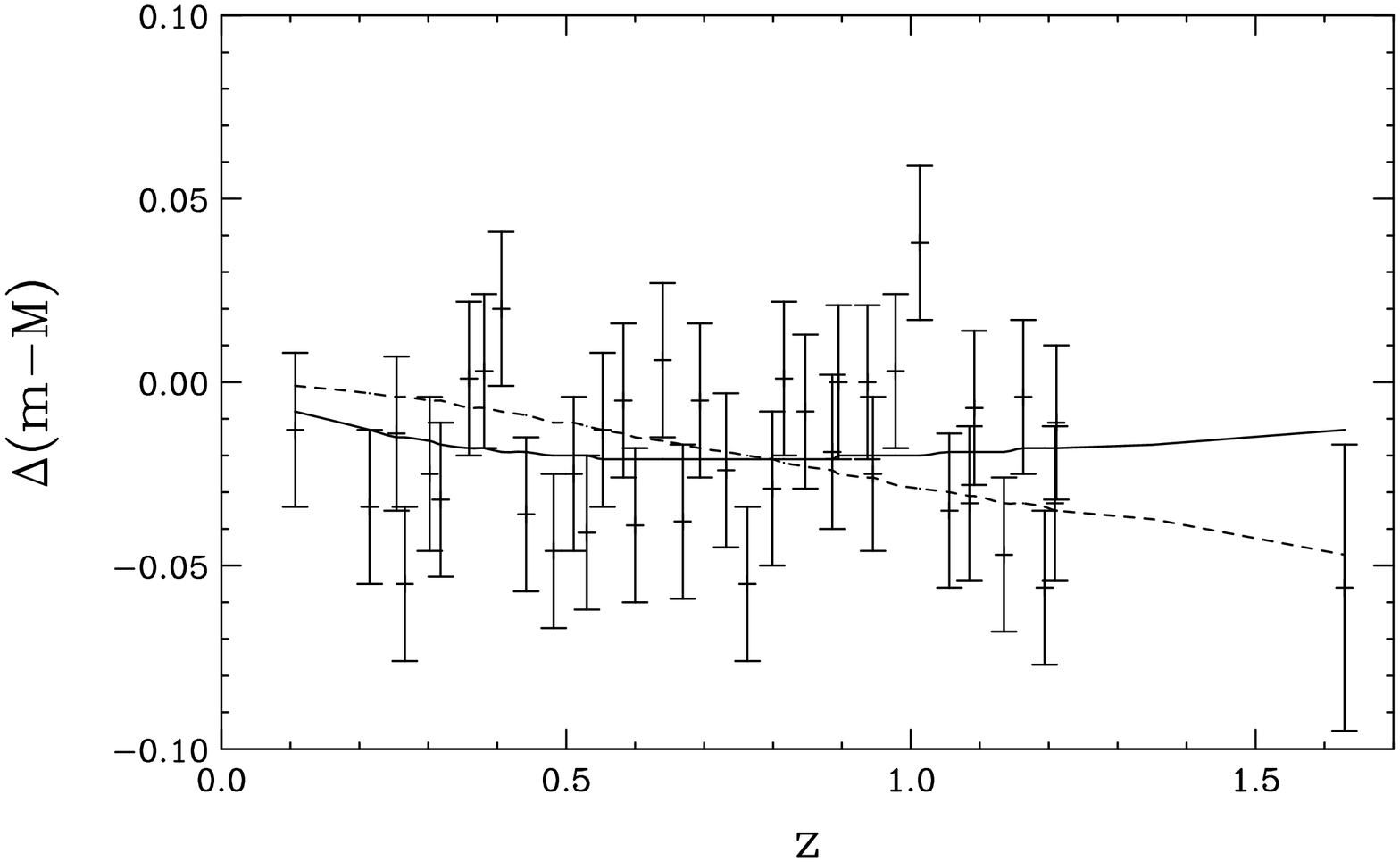,height=6cm,angle=90}
\psfig{file=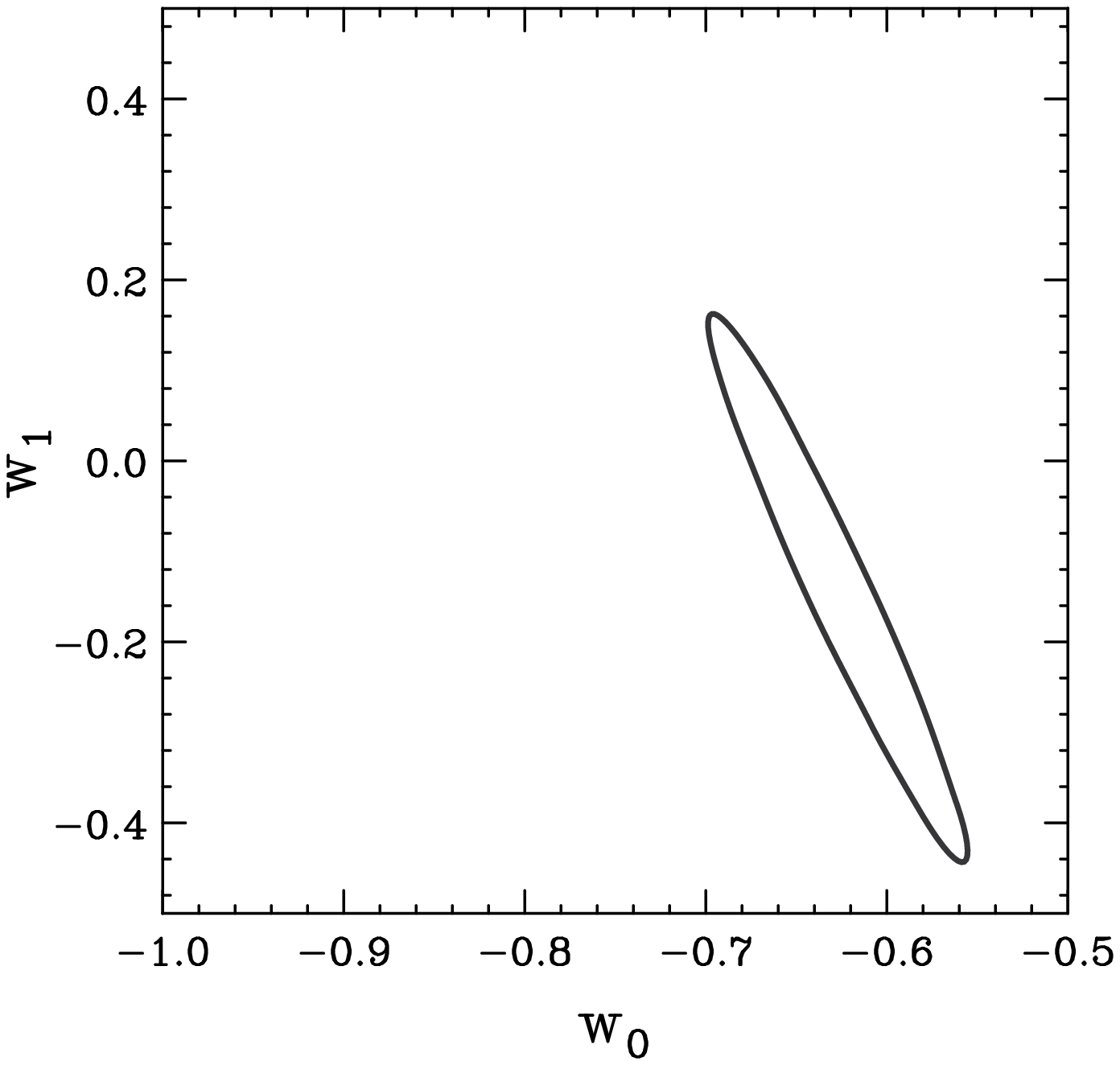,height=5.95cm,angle=90}}}
\bigskip
\caption[]{The difference in distance modulus between a quintessence 
model defined by $(\Omega_m , {w_Q}_0 , {w_Q}_1)=(0.3,-0.7,0.2)$
 and the fiducial model. 
The best fit with the assumption that $\Omega_m=0.3$ precisely is
$({w}_0 , {w}_1)=(-0.63,-0.14)$
matching a $k$-essence model. 
The 95.4\%
confidence level contour lies mainly in the lower half-plane and excludes 
the theoretical model.  }
\label{sne2}
\end{figure}

Figures ~\ref{sne1} and ~\ref{sne2} illustrate the crux of our findings that
it is also possible for SNAP to obtain datasets that give misleading evidence. 
Figure~\ref{sne1} shows data generated using a 
$k$-essence model with $(\Omega_m , {w_k}_0 , {w_k}_1)=(0.3,-0.7,-0.17)$. 
 We chose ${w_k}_1=-0.17$ so as to allow the equation of state to have
the steepest gradient (with ${w_k}_0=-0.7$) 
without violating the dominant energy
condition $|w_k|<1$ for $z<1.7$. Note that $w_k(1.7)=-0.99$.
The dashed line is the theoretical prediction and the solid line is
the best fit with  parameters $({w}_0 , {w}_1)=(-0.79,0.19)$.
In minimizing $\chi^2$ we have fixed $\Omega_m$ at 0.3. 
We see that the $k$-essence model resembles a quintessence model. 
It is noteworthy
from the $w_0-w_1$ plane in Fig.~\ref{sne1} that most of the 95.4\% 
confidence contour lies in the region $w_1>0$ even though the true model 
is $k$-essence with $w_1<0$. 
In fact, the contour entirely misses the input theoretical 
parameters \mbox{$({w_k}_0,{w_k}_1)=(-0.7,-0.17)$}.
%
%

Figure~\ref{sne2} shows results for a quintessence model that looks like a 
$k$-essence model. The theoretical parameters are 
$(\Omega_m , {w_Q}_0 , {w_Q}_1)=(0.3,-0.7,0.2)$ and the best-fit 
parameters are $({w}_0 , {w}_1)=(-0.63,-0.14)$
 keeping $\Omega_m=0.3$ during minimization. 
 The 95.4\%
confidence level excludes the theoretical model 
$({w_Q}_0,{w_Q}_1)=(-0.7,0.2)$ and lies mainly in the lower half-plane.

%
%
 
\begin{figure}[]
\centerline{\mbox{\psfig{file=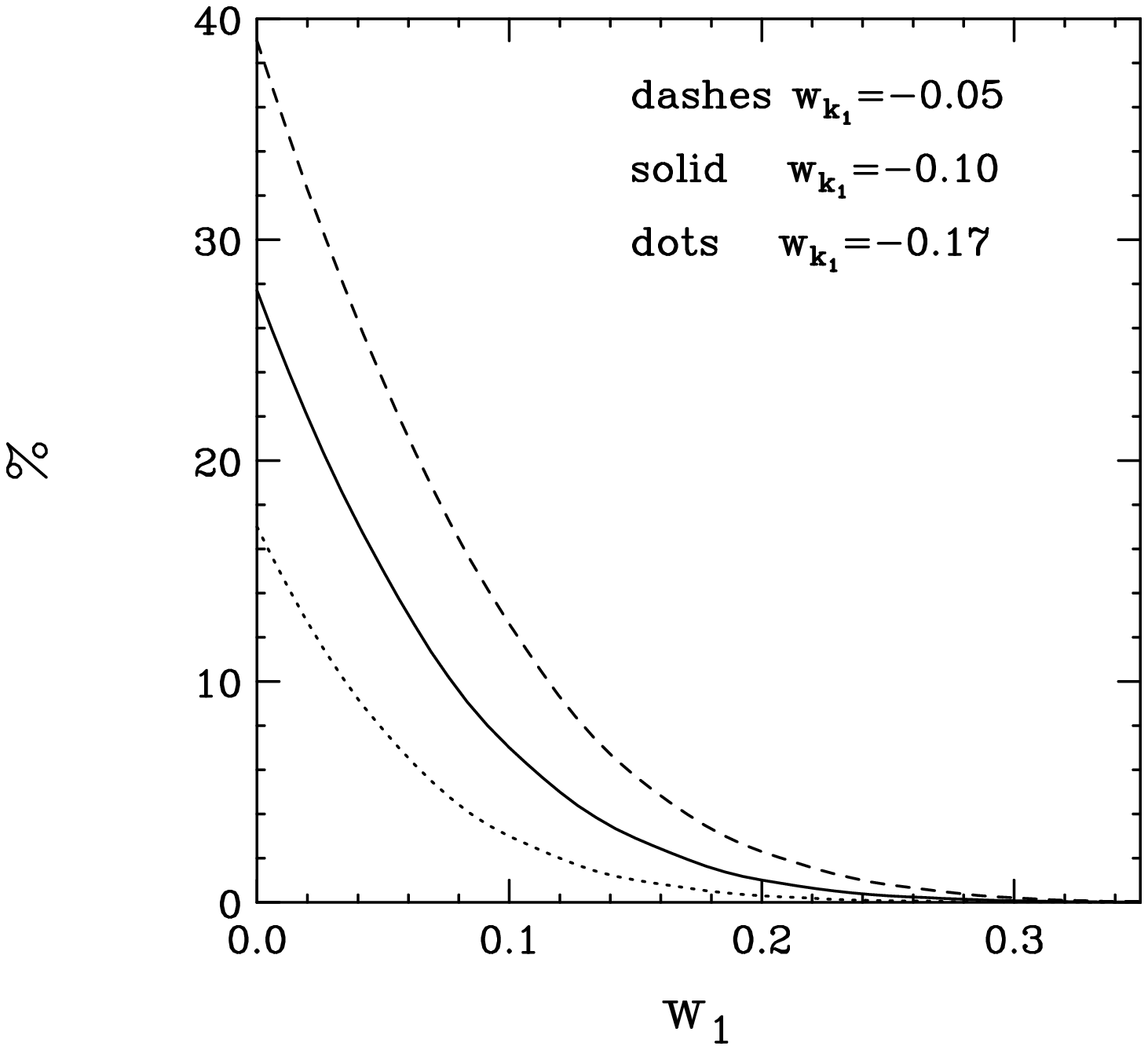,height=6cm,angle=90}
\psfig{file=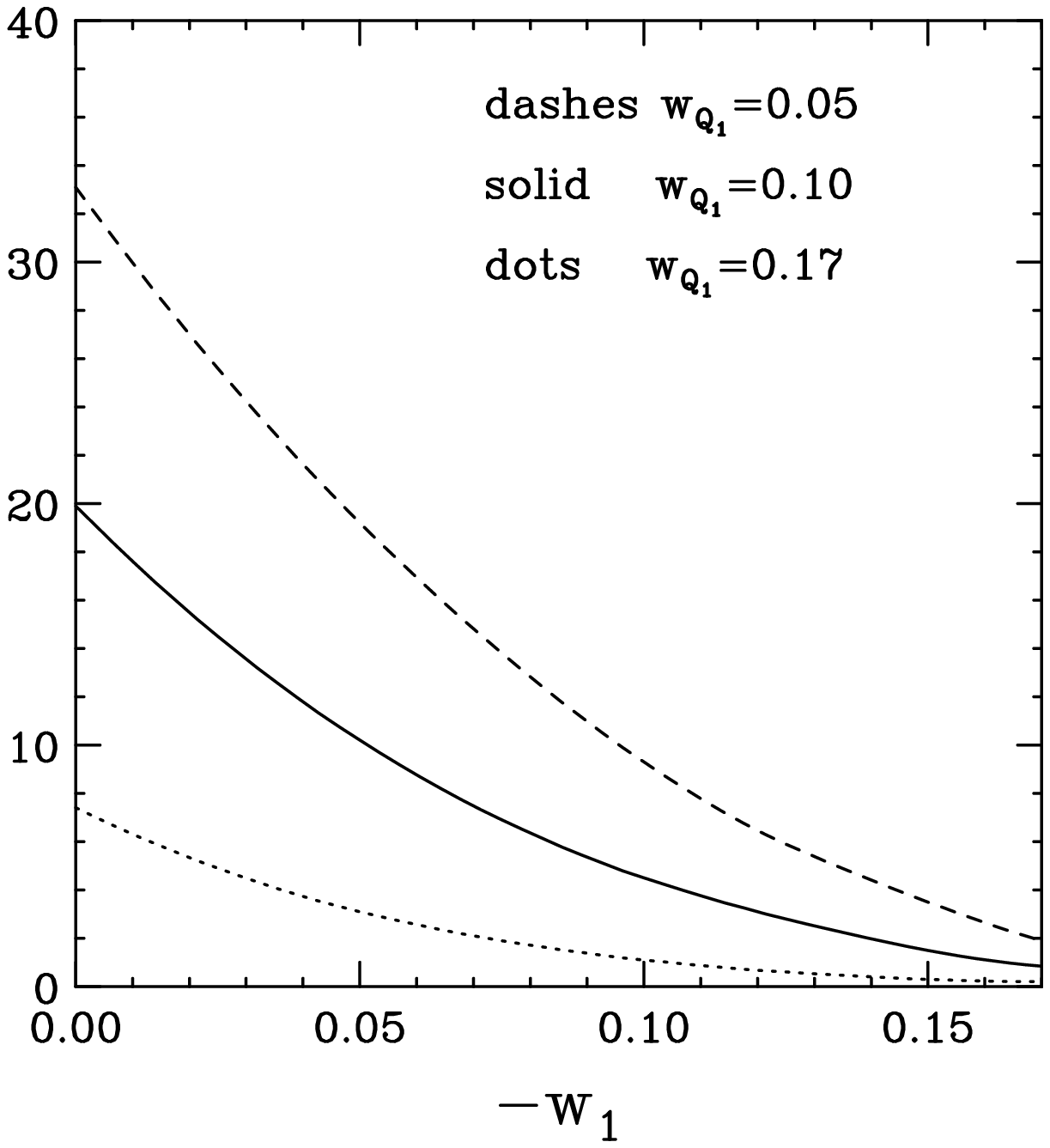,height=6cm,angle=90}}}
\bigskip
\caption[]{The percent of datasets 
derived from $k$-essence models (for different values of ${w_k}_1$) 
that have fits with a positive
$w_1$  is shown on the left. The percent of
datasets from quintessence models (for different 
values of ${w_Q}_1$) that have negative $w_1$  is shown on the right. 
Only fits
yielding $-1\leq w_0 \leq -0.6$ and $|w(z)|<1$ for $z<1.7$ are allowed. 
$\Omega_m$ is fixed to be 
$0.3$, and all theoretical models have $w(0)=-0.7$. The number of simulated 
datasets for each ${w_k}_1$ (${w_Q}_1$) is $10^5$.     }
\label{probq}
\end{figure}

Although Figs.~\ref{sne1} and \ref{sne2} present a discouraging situation,
 the probability of a dataset conspiring to create the
confusion that we have illustrated is not very large. 
In Fig.~\ref{probq} we show
the percentage of datasets simulated using a given theoretical model as 
defined by ${w_k}_1$ or ${w_Q}_1$ ($\Omega_m=0.3$ and $w(0)=-0.7$ 
for all models) that have best fits (keeping $\Omega_m=0.3$ fixed) 
that conflict with the type of input theory.
In making these plots we generated $10^5$ datasets for each $w_1$ subject 
only to the assumptions that the dominant energy condition $|w(z)|<1$ 
be satisfied 
out to redshift 1.7, and that the value of the equation of state today be 
less than $-0.6$. 
The datasets of Figs.~\ref{sne1} and \ref{sne2} 
have less than a 1\%
chance of occurring. 
 However, this 
does not mean that this sort of misrepresentation by the data
is that unlikely to occur. If the equation of state of dark energy
does not have a steep gradient, as is
necessary if the dark energy is $k$-essence and
$w(0)$ is say $-0.8$, then  to distinguish 
between the two theories becomes more difficult 
than in the examples considered. 
Moreover, additional degeneracy entering via 
uncertainties on $\Omega_m$ will contribute to confusion 
in the interpretation of SNAP data. It has been argued \cite{alb} that
the inability to discriminate between the theories is a consequence of
an inefficient fitting function, but the ambiguity is theoretical
and cannot be evaded. As shown in Ref.~\cite{maor}, $d_L(z)$ depends on $w(z)$
through a multiple-integral relation thus precluding the possibility
of a very precise measurement of $d_L$ leading to an accurate determination 
of $dw/dz$.

Even though supernova data may find it difficult to differentiate between
quintessence and $k$-essence, CMB anisotropy may be able to detect a
smoking-gun signal for $k$-essence because the speed of sound of $k$-essence 
is not unity as in quintessence models and may lead to peculiarities
in the power spectrum not considered so far \cite{k}.
\\

{\it Acknowledgments}:
This work was supported in part by a DOE grant
No. DE-FG02-95ER40896 and in part by the Wisconsin Alumni 
Research Foundation.

\end{document}